\documentclass[]{emulateapj}
\usepackage{graphicx}
\usepackage{txfonts}
\usepackage{natbib}
\newcommand{\RRgWHTband}{0.1180$\pm$0.0009}
\newcommand{\RRBVLTband}{0.1162$\pm$0.0017}
\newcommand{\RRgINTband}{0.1174$\pm$0.0017}

\shorttitle{Search for Rayleigh scattering in the atmosphere of GJ1214b}
\shortauthors{De Mooij et al.}

\begin{document}

   \title{Search for Rayleigh scattering in the atmosphere of GJ1214b}
   \author{E.J.W. de Mooij \altaffilmark{1},
           M. Brogi \altaffilmark{2},
           R.J. de Kok\altaffilmark{3},
           I.A.G Snellen\altaffilmark{2},
           B. Croll\altaffilmark{4,5},
           R. Jayawardhana\altaffilmark{1},
           H. Hoekstra\altaffilmark{2},
           G.P.P.L. Otten\altaffilmark{2},
           D.H. Bekkers\altaffilmark{2},
           S.Y. Haffert\altaffilmark{2},
           and
           J.J. van Houdt\altaffilmark{2}}
\email{demooij@astro.utoronto.ca}
\altaffiltext{1}{Department of Astronomy and Astrophysics,  University of Toronto, 50 St. George Street, Toronto, ON M5S 3H4, Canada}
\altaffiltext{2}{Leiden Observatory, Leiden University, Postbus 9513, 2300 RA, Leiden, The Netherlands;}
\altaffiltext{3}{SRON Netherlands Institute for Space Research, Sorbonnelaan 2, 3584 CA Utrecht, The Netherlands;}
\altaffiltext{4}{Kavli Institute for Astrophysics and Space Research, Massachusetts Institute of Technology, Cambridge, MA 02139;}
\altaffiltext{5}{NASA Sagan Fellow}

\begin{abstract}
We investigate the atmosphere of GJ1214b, a transiting super-Earth planet with a low mean density, by measuring its transit depth as a function of wavelength in the blue optical portion of the spectrum. It is thought that this planet is either a mini-Neptune, consisting of a rocky core with a thick, hydrogen-rich atmosphere, or a planet with a composition dominated by water. 
 Most observations favor a water-dominated atmosphere with a small scale-height, however, some observations indicate that GJ1214b could have an extended atmosphere with a cloud layer muting the molecular features. In an atmosphere with a large scale-height, Rayleigh scattering at blue wavelengths is likely to cause a measurable increase in the apparent size of the planet towards the blue.
 We observed the transit of GJ1214b in the B-band with the FOcal Reducing Spectrograph (FORS) at the Very Large Telescope (VLT) and in the g-band with both ACAM on the William Hershel Telescope (WHT) and the Wide Field Camera (WFC) at the Isaac Newton Telescope (INT). We find a planet-to-star radius ratio in the B-band of \RRBVLTband, and in the g-band \RRgWHTband\ and \RRgINTband\ for the WHT \& INT observations respectively. These optical data do not show significant deviations from previous measurements at longer wavelengths. In fact, a flat transmission spectrum across all wavelengths best describes the combined observations. When atmospheric models are considered a small scale-height water-dominated model fits the data best.
\end{abstract}


\section{Introduction}

The transiting planet GJ1214b \citep{charbonneauetal09} is currently the best studied super-Earth. With a radius of 2.6 R$_{Earth}$ and a mass of 6.5 M$_{Earth}$ its density is significantly lower than that of the Earth. \cite{rogersetal10} explain the low density of GJ1214b in the context of three different formation models; two of these models result in a thick atmosphere with a low mean-molecular weight and a large scale-height, while the third scenario, that GJ1214 is a water-world, predicts a small scale-height due to the large mean-molecular weight of water compared to hydrogen. 

Models for GJ1214b's atmosphere showed that in the case of a hydrogen-rich atmosphere, signatures of the atmosphere should be well within reach of current instrumentation \citep[][]{millerricciandfortney10,howeandburrows12}. Many observations probing the atmosphere of GJ1214 have been presented in the literature \citep[][]{beanetal10,beanetal11,crolletal11,crossfieldetal11,bertaetal12,demooijetal12,murgasetal12,naritaetal12,fraineetal13,teskeetal13}. Although most of these observations did not show the expected features for an extended atmosphere, \cite{crolletal11} found  an increased radius in the K-band indicative of molecular absorption, and \cite{demooijetal12}, hereafter Paper I, reported a tentative signal from Rayleigh scattering towards blue wavelengths. In the case that the atmosphere of GJ1214b is extended, the lack of spectral features at longer wavelengths can be explained by a low abundance of methane, as well as a cloud-layer that masks the features.

Here we present the results of transit observations at blue optical wavelengths in order to investigate the tentative signal of Rayleigh scattering found in Paper I. In Sec.~\ref{sec:obs} we report the observations. In Sec.~\ref{sec:dr} we describe the data-reduction, corrections for systematic effects and the transit fitting. In Sec.~\ref{sec:discuss} we discuss our results in the context of atmospheric models. Finally, we present our conclusion in Sec.~\ref{sec:concl}

\section{Observations}\label{sec:obs}

\begin{figure}
\centering
\includegraphics[width=8cm]{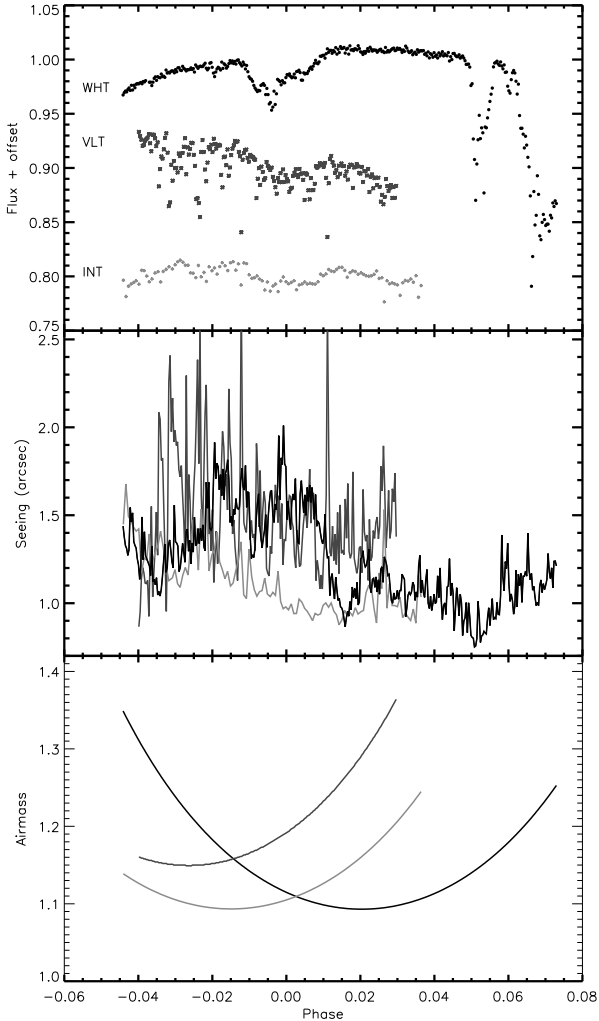}
\caption{ Top panel: Lightcurves of GJ1214 without the use of a reference star for the three nights of observation. Middle panel: The seeing during the three nights, the WHT, VLT and INT observations are shown with black, dark grey and light grey lines respectively. Bottom panel: The geometric airmass for the three transit observations.}
\label{fig:obs_details}
\end{figure}

\subsection{Transit observations with ACAM}
On the night of May 22, 2011, a transit of GJ1214b was observed in the g-band ($\lambda_c$=4996$\AA$) with ACAM \citep{ACAM08} at the William Herschel Telescope. The observations started at 23:59 UT, and lasted for $\sim 4.4$ hours. An exposure time of 30 seconds was used, resulting in a total of 364 frames, with an average cadence of 44 seconds. The circular field of view of ACAM is 8 arcmin in diameter, allowing the observation of several reference stars simultaneous with the target. Due to the relative faintness of GJ1214 in g-band, we did not defocus the telescope. Despite guiding, flexing between the instrument and the guider resulted in a drift of 10 pixels (2.5'') during the night. Towards the end of the night the sky became non-photometric, with an extinction of up to 20\%, while the median seeing was 1.2'' (see Fig.~\ref{fig:obs_details}).

\subsection{Transit observations with FORS}
A second transit of GJ1214 was observed with the FOcal Reducing Spectrograph (FORS) on the Very Large Telescope (VLT) on July 12, 2011 in the B-high filter ($\lambda_c$=4543$\AA$). The observations, which lasted for 2h40m, started at 01:28UT. A total of 205 frames with an exposure time of 20 sec were obtained, with an average cadence of 47 seconds. As with the WHT observations, the telescope was not defocused. To increase the cadence the 2x2 binning mode was used, yielding a pixel scale of 0.25''/pixel. Although FORS consists of 2 detectors, only data from one detector was used, due to a systematic difference in the photometry between the detectors. The detector used has an unvignetted field of view of 7' by 4'. The seeing during the observations was highly variable, varying between 1'' and 3'' with a median of 1.4'' (see  Fig.~\ref{fig:obs_details}).

\subsection{Transit observations with WFC}
On the night of May 6, 2012, a transit of GJ1214b was observed in the g-band ($\lambda_c$=4962$\AA$) with the Wide Field Camera (WFC) of the Isaac Newton Telescope (INT). The observations lasted 3 hours and started at 02:35UT. During this time 113 frames were obtained with an average cadence of 97 seconds and an exposure time of 60 seconds. Although the camera consists of four CCDs, each with a pixel scale of 0.33''/pixel, only the central CCD (CCD4) was used, with a field-of-view of $\sim$22.7' by 11.4'. The telescope was not defocused for these observations.The atmosphere was stable during the observations, and the median seeing was 1.4'' (see  Fig.~\ref{fig:obs_details}).

\section{Data reduction \& Analysis}~\label{sec:dr}

\begin{figure*}
\centering
\includegraphics[width=16cm]{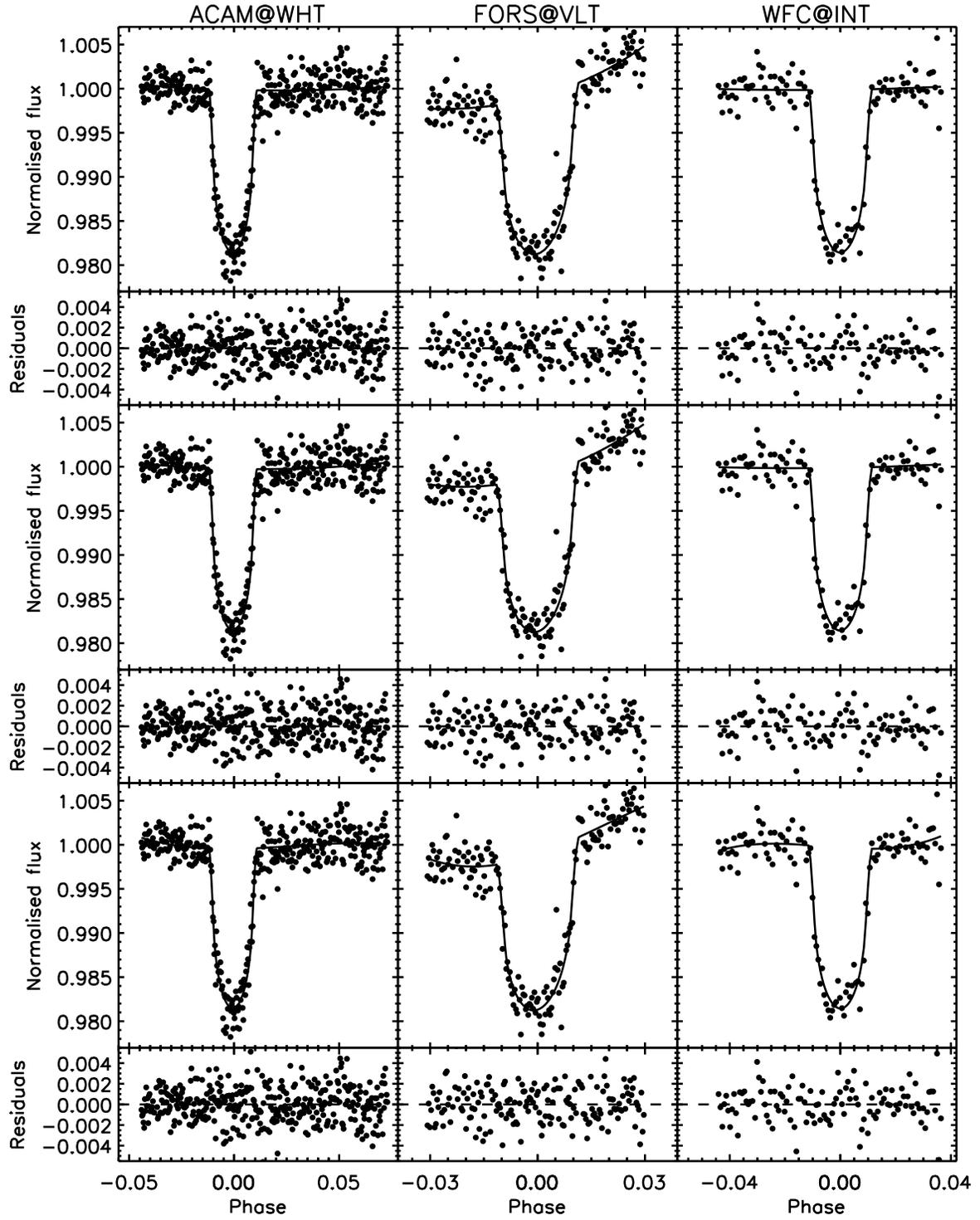}
\caption{Light curves \& residuals for the WHT (left panels), VLT (middle panels) and INT (right panels) datasets. Overplotted are the best fitting models for a baseline based on airmass (first two rows), second order polynomial (middle two rows) and third order polynomial (bottom two rows).}
\label{fig:lcs}
\end{figure*}

The data from all telescopes were reduced in similar ways: after bias subtraction the data were flatfielded using a flatfield created from twilight flats. The atmospheric dispersion corrector (ADC) of the FORS instrument is known to introduce rotator angle dependent variations in the flatfield \citep{moehleretal10}. Therefore flatfields for the B$_{high}$ filter at different rotator angles were obtained from the ESO archive for the period surrounding our observations. These flatfields were combined into a single master flat. 

Subsequently, aperture photometry was performed on GJ1214 and a set of reference stars. An aperture radius of 18, 13 and 13 pixels was used for the ACAM, FORS and WFC observations respectively (4.5, 3.75 and 4.3 arcsec). These apertures were chosen to minimise the noise in the respective light curves. The photometry from multiple reference stars was combined into a master reference light curve for each of the observations. For the INT observations 13 reference stars were used, while the limited field-of-view for the other instruments limited the number of useable reference stars to 4 and 3 for the WHT and VLT respectively. Although there are more stars inside the field of view, many of them are either saturated or show strong systematic effects (for the VLT this might be due to the rotator angle dependence of the flat field). Subsequently the light curve for GJ1214 was divided by the master reference light curve and normalised. 

\subsection{Light curve fitting}
The normalised light curves are shown in Fig.~\ref{fig:lcs}. No strong systematic effects are seen in the WHT and INT light curves, while the VLT light curve shows a smooth gradient. We fit for this trend simultaneous with the transit depth using a Markov-Chain Monte-Carlo (MCMC) method. The transit is modeled using the \cite{mandelandagol02} routines with the scaled semi-major axis and the impact parameters fixed to those from \cite{beanetal10} (a/R$_*$=14.9749 and b=0.2779), the limb-darkening was also kept fixed, for the VLT observations to the values from \cite{claret00}, while for the WHT \& INT observations limb-darkening coefficients from \cite{claret04} were used. We used the quadratic limb darkening coefficients for a star with T$_{eff}$=3030K and log(g)=5.0. The time of mid-transit and the planet-to-star radius ratio are free parameters. We verified that when we include the impact parameter, semi-major axis, and limb-darkening as free parameters, we obtain consistent results.

Three different models were used to correct the systematic effects in the baseline:  a model depending on the geometric airmass, a second-order polynomial baseline model and a third-order polynomial baseline model. For each combination of light curve and baseline model we created five separate MCMC chains of 200,000 steps for which the first 20,000 steps were discarded to remove the burn-in. Subsequently these five chains were merged, after verifying that they had converged using the Gelman-Rubin statistic \citep{gelmanandrubin92}. The results of these MCMC chains are given in Table~\ref{tab:rprs_baselines}. 

The impact of red-noise was assessed using the residual permutation (``Prayer-bead'') method \citep[e.g.][]{gillonetal07} for each of the datasets. The 16\% and 84\% distribution of the fitted parameters are taken as the $\pm$ 1$\sigma$ confidence interval. The radius-ratios with uncertainties as determined from this analysis are given in Table~\ref{tab:rprs_prayer_bead}. As can be seen the uncertainties from this analysis indicate that the contribution from red-noise is small. We therefore adopt the measurements from the MCMC analysis for the rest of this work. 

\begin{figure*}
\centering
\includegraphics[width=16cm]{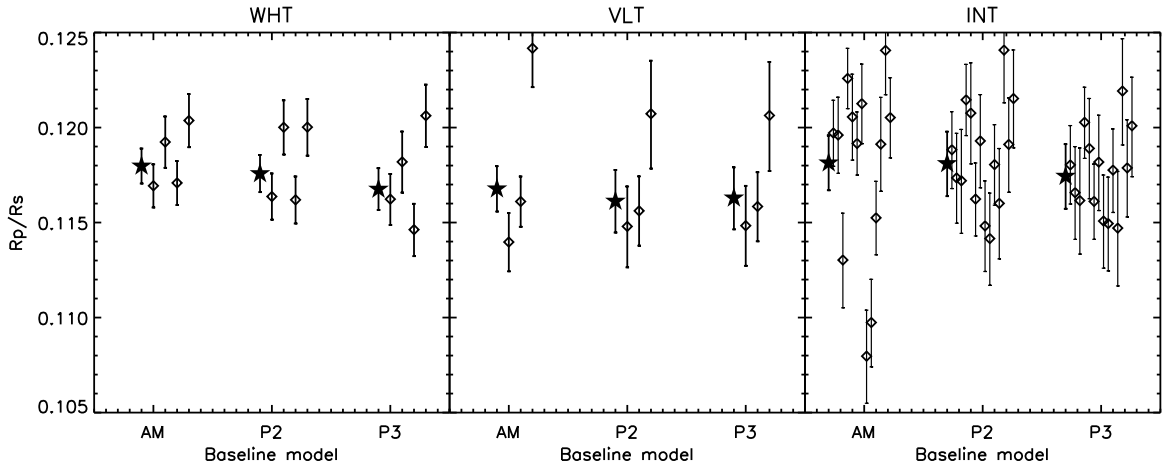}
\caption{Measured planet-to-star radius ratios for the three datasets using three different baseline corrections. The filled stars are the values determined for the master lightcurve (the light curve of the target divided by the combined light curve of the reference stars), while the open diamonds are the radius ratios as determined when light curve of the target is correct with the individual reference stars. The three different baselines are: Airmass (AM), 2nd order polynomial (P2) and third order polynomial (P3).}
\label{fig:rprs}
\end{figure*}

 As an additional check on the robustness of the fit we determined the radius-ratio for light curves generated using individual reference stars. In Fig.~\ref{fig:rprs} we show for each of the three datasets the measured radius ratios for both a reference signal based on an ensemble of light curves as well as for the individual light curves. As can be seen most of the measurements for the individual stars agree well with the measurement for the combined reference signal, except for the airmass baseline fit to the INT g-band data. For the rest of the paper we adopt the baseline with airmass only for the WHT observations, the results for the second-order polynomial for the VLT observations and the model with the third order polynomial baseline for the INT observations. The RMS of the residuals of the data with the selected baselines are  1.68 mmag, 1.89 mmag, and 1.66 mmag for the WHT, VLT and INT data respectively

\begin{table}
\caption{Measurements of R$_p$/R$_*$ for the three observations and different baseline models.}
\label{tab:rprs_baselines}
\centering
\small

\renewcommand{\arraystretch}{1.35}
\begin{tabular}{l c c c}
\hline\hline
  Baseline  & WHT g-band & VLT B-band & INT g-band \\
\hline
 Airmass       & 0.1180$\pm$0.0009 & 0.1168$\pm$0.0012 & 0.1181$\pm$ 0.0014 \\
 2$^{nd}$ order & 0.1176$\pm$0.0010 & 0.1161$\pm$0.0017 & 0.1181$\pm$ 0.0017 \\
 3$^{rd}$ order & 0.1168$\pm$0.0011 & 0.1163$\pm$0.0017 & 0.1174$\pm$ 0.0017 \\
\hline
\end{tabular}
\end{table}

\begin{table}
\caption{Measurements of R$_p$/R$_*$ for the three observations and different baseline models from the residual permutation (prayer-bead) method. The quoted R$_p$/R$_*$ is for the median of the distribution and the upper and lower uncertainties are for the 16\% \& 84\% levels respectively.}
\label{tab:rprs_prayer_bead}
\centering
\renewcommand{\arraystretch}{1.35}
\begin{tabular}{l c c c}
\hline\hline
 Basline  & WHT g-band & VLT B-band & INT g-band \\
\hline
 Airmass       & 0.1181$^{+0.0005}_{-0.0010}$ & 0.1165$^{+0.0010}_{-0.0003}$ & 0.1180$^{+0.0008}_{-0.0009}$\\
 2$^{nd}$ order & 0.1174$^{+0.0006}_{-0.0009}$ & 0.1160$^{+0.0006}_{-0.0003}$ & 0.1180 $^{+0.0008}_{-0.0009}$ \\
 3$^{rd}$ order & 0.1165$^{+0.0006}_{-0.0007}$ & 0.1162$^{+0.0002}_{-0.0003}$ & 0.1172$^{+0.0009}_{-0.0008}$ \\
\hline
\end{tabular}
\end{table}

\section{Results \& Discussion}\label{sec:discuss}
\begin{figure*}
\centering
\includegraphics[width=16cm]{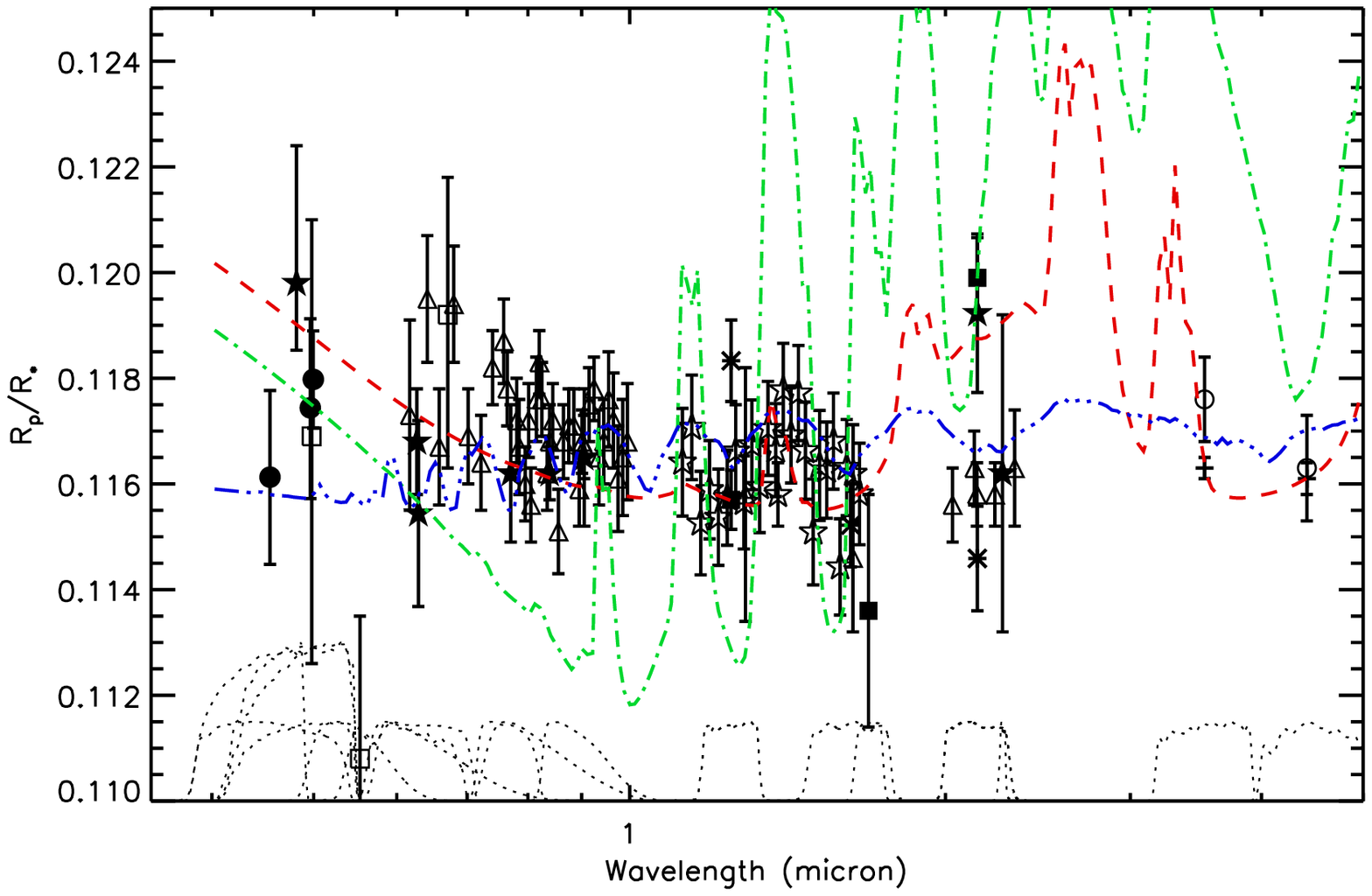}
\caption{Transmission spectrum of GJ1214b, including all available measurements from the literature. The filled circles are the observations from this work, the filled stars are for data from Paper I, the filled squares are for data from \cite{crolletal11}, the open triangles  are for data from \cite{beanetal11}, the open stars are for data from \cite{bertaetal12}, the open circles are for data from \cite{desertetal11}, the asterixes are for data from \cite{naritaetal12}, the crosses are for data from \cite{fraineetal13}, and the open squares are for data from \cite{teskeetal13}. Overplotted are three models for the atmosphere of GJ1214b: a hydrogen dominated atmosphere with solar composition (green line), a hydrogen dominated atmosphere with clouds and low methane abundance (red line) and a water dominated atmosphere (blue line). The dashed curves at the bottom are the transmission curves for the different filters. Note that for clarity the filter curves for the data from \cite{beanetal11} and \cite{bertaetal12} have been omited, since they are effectively flat.}
\label{fig:spec_all}
\end{figure*}

We find planet-to-star radius ratios of \RRgWHTband, \RRBVLTband\ and \RRgINTband\ respectively for the WHT g-band, VLT B-band and INT g-band. These radii are consistent with radius-ratios measured at longer wavelength \citep[e.g. from the MEarth data][]{charbonneauetal09}, and do not confirm the potential 2$\sigma$ deviation in g-band found in Paper I. In addition to the data above,  GJ1214b has been observed by many other groups, and the available data from the literature spans wavelengths from $\sim$0.4 $\mu$m to $\sim$ 5 $\mu$m \citep[][Paper I]{beanetal10,beanetal11,desertetal11,crolletal11,bertaetal12,naritaetal12,fraineetal13,teskeetal13}. Although most of these datasets use the same parameters for the system \citep[from][]{beanetal10}, some \citep[e.g.][]{bertaetal11} use a different set of parameters, which could cause an offset, making a direct comparison between the measured radius-ratios more difficult. In addition, we note that GJ1214 is an active star, and that the presence of starspots could bias the measurement of the radius measurements (see Sect.~\ref{sec:starspots}). Since the measurements from the literature were made at different dates, and therefore different levels of stellar activity, this could influence the observed transmission spectrum.  In Fig.~\ref{fig:spec_all} we show the transmission spectrum of GJ1214b based on our measurements and those from the literature listed above\footnote{Note that in the case of papers with multiple measurements with the same wavelength and instrument we use the value for the combined fit to the data.}

\subsection{Comparison with atmospheric models}

We compared the observed wavelength dependent transmission spectrum with a set of models for three different scenarios for the nature of the atmosphere of GJ1214: an atmosphere with a small scale-height (a ``water world''), a large scale-height atmosphere without clouds and an atmosphere with clouds and a large scale-height. The models are the same as used in Paper I. 

\begin{figure*}
\centering
\includegraphics[width=16cm]{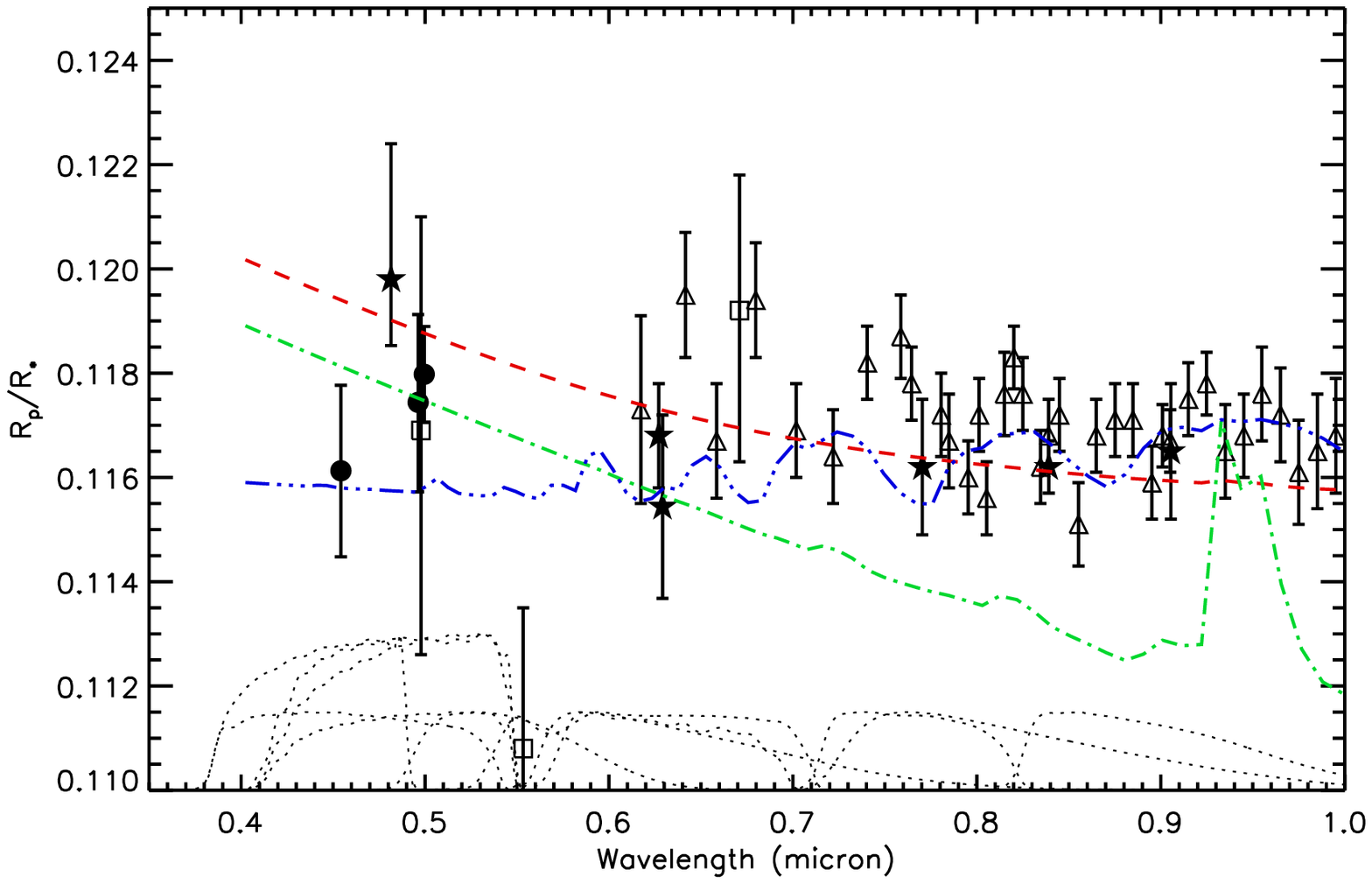}
\caption{Same as Fig.~\ref{fig:spec_all} but now only focusing on the optical part of the spectrum.}
\label{fig:spec_opt}
\end{figure*}

We have overplotted the models on the data in Figs.~\ref{fig:spec_all}~and~\ref{fig:spec_opt}. As can be seen the model with the large scale-height and no clouds (green  dashed-dotted line) can clearly be excluded. The water-world model (blue, dashed-triple dotted line) and the model with a large scale-height and a cloud layer (red dashed line) are both consistent with the data. 

Assuming that all the uncertainties on the measurements are reliable, we can calculate the formal $\chi^2$ for each of the three models. For the water-world $\chi^2$=190, for the cloudless atmosphere $\chi^2$=4003 and for the atmosphere with clouds $\chi^2$=289, while a flat line gives $\chi^2$=148. From this it follows that the current available data favour a flat transmission spectrum (i.e. no atmosphere or a grey atmosphere), followed by a water dominated atmosphere. In this paper a total of 92 measurements of the radius are used. Note that assigning degrees of freedom to each of these models is not very useful, since there are many different parameters that can be adjusted (e.g. composition and temperature-pressure profile). 

\subsection{Potential origins of a flat transmission spectrum}\label{sec:flat}
\begin{figure}
\centering
\vspace{0.7cm}
\includegraphics[width=8cm]{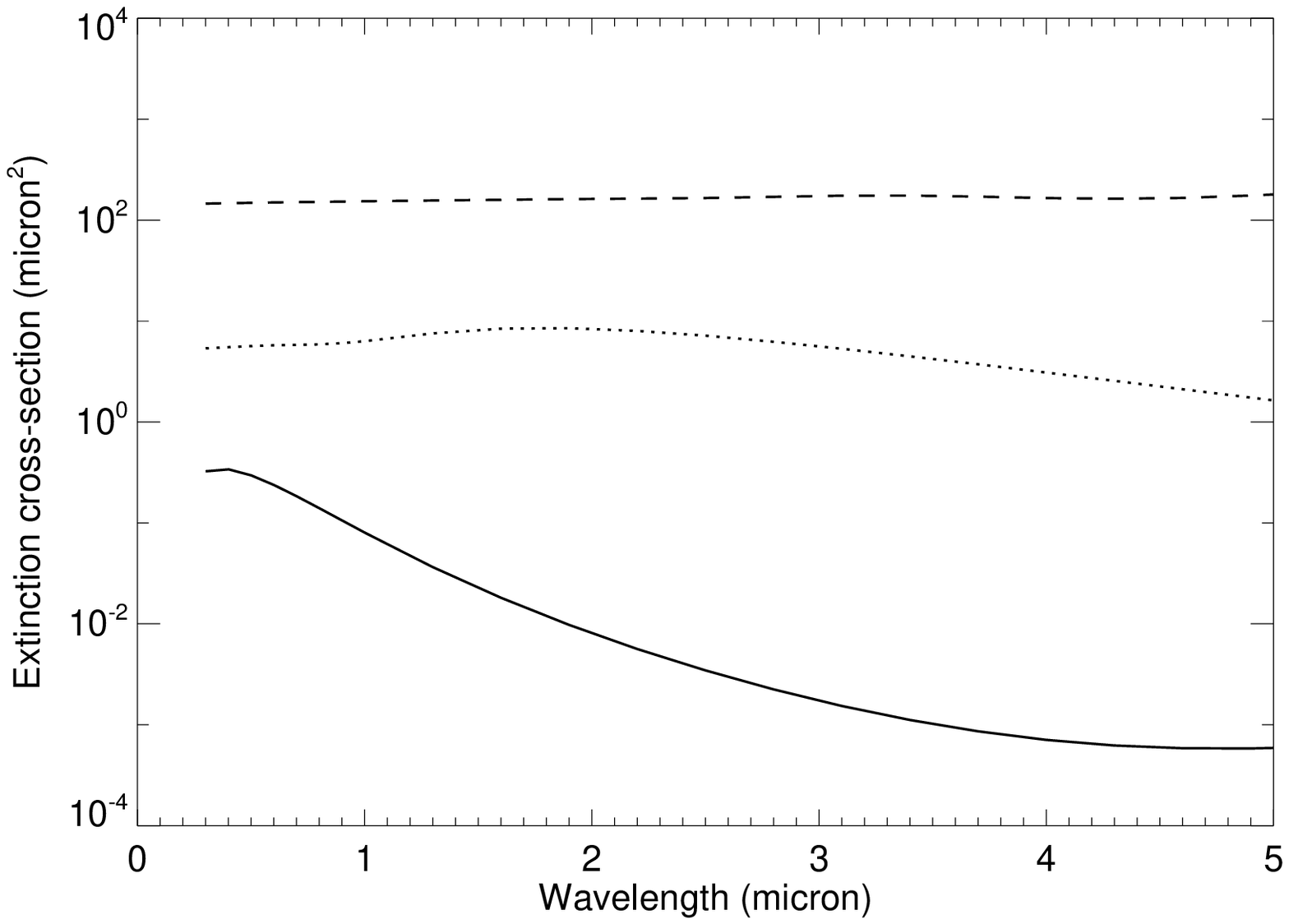}
\caption{Extinction cross-section for Mg$_2$SiO$_4$ particles with narrow log-normal size distributions around 5.0, 1.0 and 0.2$\mu$m respectively from top to bottom.}
\label{fig:mie_mg2sio4}
\end{figure}
\begin{figure}
\vspace{0.7cm}
\centering
\includegraphics[width=8cm]{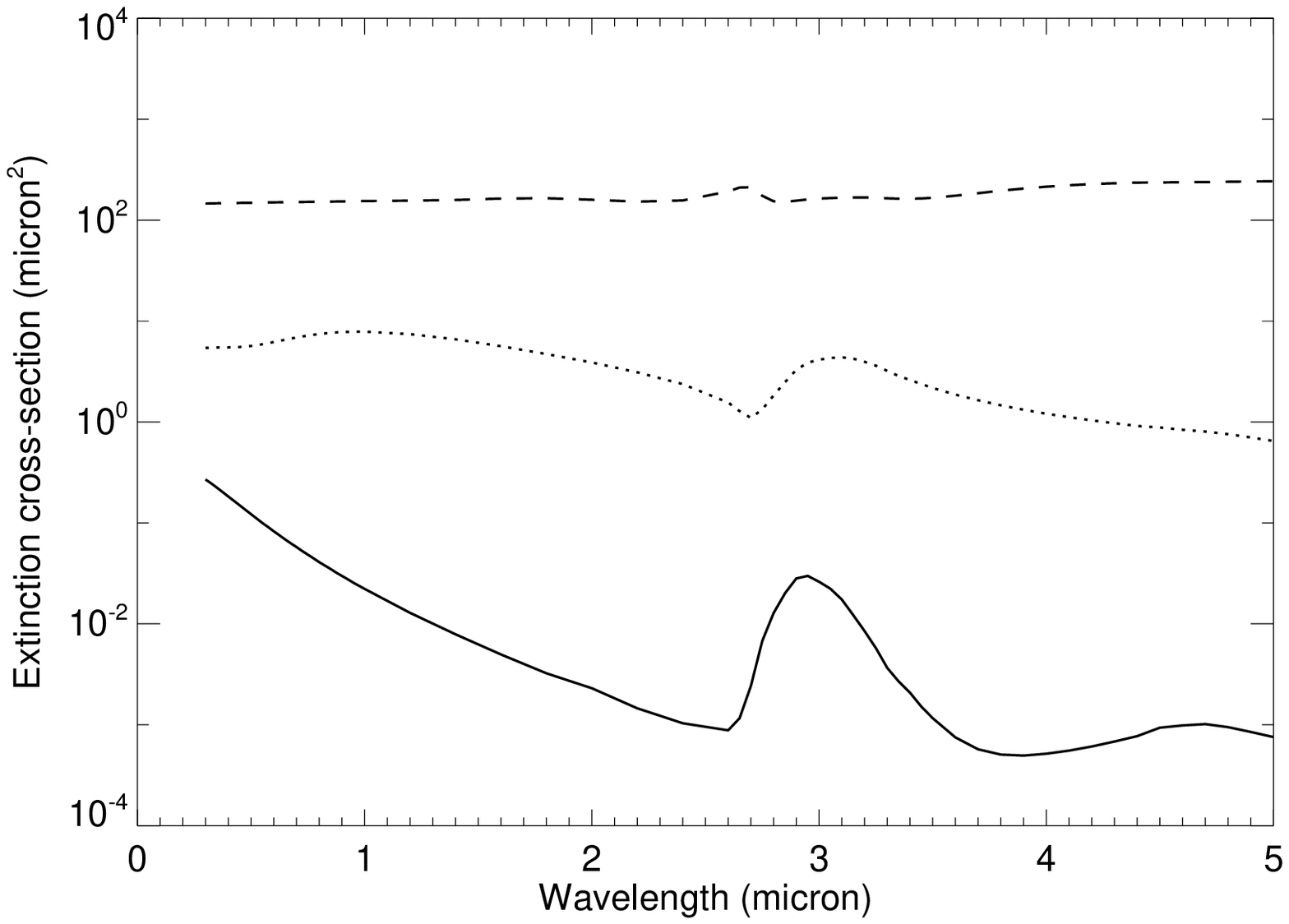}
\caption{Extinction cross-section for H$_2$O droplets with narrow log normal size distributions around 5.0, 1.0 and 0.2$\mu$m respectively from top to bottom.}
\label{fig:mie_h2o}
\end{figure}
As discussed above, and by various other authors, the best fit to the collective set
of measurements of the transit depth is a flat line. Here we briefly discuss some
physical possibilities that can create such a flat transmission spectrum. Although unlikely, a flat transmission spectrum could indicate that GJ1214b lacks an atmosphere with the density of the planet, which requires either a water-dominated planet or a planet with an extended envelope~\citep{rogersetal10}. 

In the case of an extended atmosphere, the presence of clouds can also result in a grey transmission spectrum. This point is mentioned by several authors in a qualitative manner \citep[e.g.][]{beanetal10, demooijetal12, howeandburrows12}, but it needs pointing out that such a cloud would have to meet specific requirements to produce a flat spectrum \citep[see also][]{bertaetal12}. One option is that there is an optically thick cloud, which itself has a wavelength-independent extinction, at high enough altitudes to cover any gas absorption features. Such a cloud can be vertically extended, as long as it is optically thick when seen in transmission. Because of the wavelength-independent extinction, the altitude at which the cloud becomes optically thick will also be independent of wavelength. For cloud properties to be constant with wavelength, the particles need to be larger in size than the wavelength range. Fig.~\ref{fig:mie_mg2sio4} shows an example of the extinction cross-sections as a function of wavelength for Mg$_2$SiO$_4$ particles, which has only slowly varying optical constants along these wavelengths and is not strongly absorbing. On the other hand, Fig.~\ref{fig:mie_h2o} shows the extinction cross-sections for the same particle size distributions for water droplets, which has a clear absorption feature around 3 micron. Even for the large particles, the extinction cross-section is significantly larger there than at other wavelengths. For both the Mg$_2$SiO$_4$ and H$_2$O the cross-sections are calculated using a Mie code based on \cite{derooijandvanderstap84}.

Alternatively, if the clouds have a very sharp decrease of optical depth above the cloud tops, that could also result in a featureless transmission spectrum. If the cloud is optically thick along the slant path through the atmosphere, variations of cloud properties with wavelength are irrelevant. The cloud then still needs to be above the altitude where gas absorption is important. Note that at present we cannot constrain a pressure range of the cloud tops, since we do not know the gas abundances. For instance, \cite{lammeretal13} show that atmospheric escape can be very important for GJ 1214b, which will change the composition of the atmosphere. The pressure limits mentioned by \cite{bertaetal12} are only valid for a solar composition atmosphere in chemical equilibrium.

Finaly, a sharp drop of gas volume mixing ratio, for instance due to condensation, photolysis or ionization \citep[e.g.][]{millerriccikempton12}, can make a transmission spectrum appear flatter across all wavelengths, although this is unlikely to yield a perfectly flat transmission spectrum.

\subsection{Stellar variability}\label{sec:starspots}
\begin{figure}
\centering
\includegraphics[width=8cm]{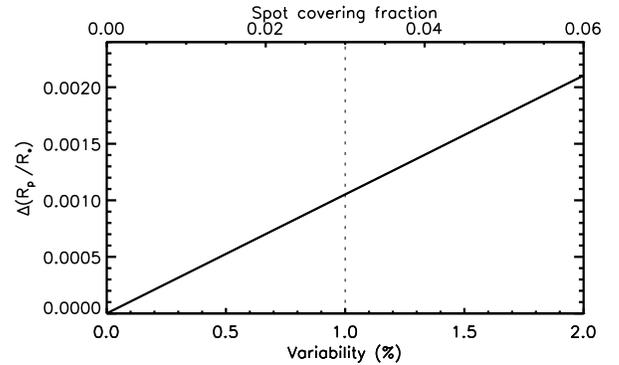}
\caption{The change in R$_p$/R$_*$ in the g-band as a function of stellar variability in the I-band. The corresponding spot coverage fraction, assuming a spot temperature of 2800K is indicated above. The stellar variability as measured by \cite{bertaetal11} is $\sim$1\% in the the MEarth bandpass, comparable to the I-band used here, and is indicated by a dashed line.}
\label{fig:spotfrac}
\end{figure}
In the discovery paper~\cite{charbonneauetal09} noted that host-star, GJ1214, is variable with a period of tens of days, which they attribute to starspots rotating in and out of view. \cite{bertaetal11} used more measurements from MEarth, and found an average variability of 1\% on a period of $\sim$50 days. Unocculted starspots, whether contributing to the stellar variability or not, can, as discussed in Paper I  \citep[see also][]{singetal11b,desertetal09}, have a significant impact on the observed transmission spectrum. The WHT, INT and VLT observations were taken over the span of approximately a year, and are, most-likely, at different levels of stellar variability. This could lead to systematic differences in the measured radius ratios. Assuming a spot temperature of 2800K ($\sim$200K lower than the temperature of the stellar photosphere), the $\sim$1\% variability from \cite{bertaetal11} corresponds to a bias in the radius ratio of $\sim$0.001 in the g-band(see Fig.~\ref{fig:spotfrac}). Since we do not have contemporary measurements of the variability for all the observations, we do not correct our measurements for this effect.

As noted in Paper I, a background of unocculted starspots that does not contribute to the stellar activity, could also significantly alter the shape of the transmission spectrum.

\section{Conclusions}\label{sec:concl}
We have observed three transits of GJ1214b at blue optical wavelengths in order to search for the signature of Rayleigh scattering in its atmosphere, which should be present if its atmosphere has a low mean molecular weight and no large cloud particles. We find planet-to-star radius ratios of \RRgWHTband, \RRBVLTband\ and \RRgINTband\ in the g$_{WHT}$-, B$_{high}$- and g$_{INT}$-bands respectively, and we do not confirm the potential 2$\sigma$ signal of Rayleigh scattering presented in Paper I. Combining these measurements with values from the literature, and comparing the full wavelength dependent transmission spectrum with atmospheric models we find that a model with a large scale height and no cloud layer is incompatible with the available data. Both a model with a large scale height and a cloud layer and a model for an atmosphere with a small scale-height are a better fit to the data with the water model favoured over the model for an extended atmosphere with clouds. We therefore conclude that, based on the currently available data, GJ1214b probably has an atmosphere with a small scale-height, although it should be noted that a featureless transmission spectrum provides the lowest $\chi^2$. This flat spectrum could be due to clouds, as discussed in Sect.~\ref{sec:flat}.

\begin{acknowledgements}
We are grateful to the staff of the Isaac Newton Group, ESO's Paranal observatory. The Isaac Newton Telescope is operated on the island of La Palma by the Isaac Newton Group in the Spanish Observatorio del Roque de los Muchachos of the Instituto de Astrof\'isica de Canarias. Based on observations collected at the European Southern Observatory, Chile (287.C-5035). E.dM. is supported in part by an Ontario Postdoctoral Fellowship. This work is supported in large part by grants to R.J. from the Natural Sciences and Engineering Research Council of Canada, which also funds B.C.'s research. B.C.’s work was performed under contract with the California Institute of Technology funded by NASA through the Sagan Fellowship Program.
\end{acknowledgements}

\end{document}